# High pressure operation of the triple-GEM detector in pure Ne, Ar and Xe

A. Bondar, A. Buzulutskov [*], L. Shekhtman

*Budker Institute of Nuclear Physics, 630090 Novosibirsk, Russia*

**Abstract**

We study the performance of the triple-GEM (Gas Electron Multiplier) detector in pure noble gases Ne, Ar and Xe, at different pressures varying from 1 to 10 atm. In Ar and Xe, the maximum attainable gain of the detector abruptly drops down for pressures exceeding 3 atm. In contrast, the maximum gain in Ne was found to increase with pressure, reaching a value of $10^5$ at 7 atm. The results obtained are of particular interest for developing noble gas-based cryogenic particle detectors for solar neutrino and dark matter search.

---

[*] Corresponding author. Tel: +7-3832-302024; fax: +7-3832-342163.
Email: buzulu@inp.nsk.su

This study is motivated by the growing interest in developing cryogenic double-phase particle detectors for solar neutrino and dark matter search [1,2]. In such detectors, the ionization produced in a noble liquid by a neutrino or weakly ionizing particle interaction, in Ne or Xe correspondingly, is extracted from the liquid to a gas phase, where it is detected with the help of the gas multiplier.

In traditional gaseous detectors, namely in the multi-wire proportional and parallel-plate avalanche chambers, the maximum gain obtained in pure noble gases is by far too low due to photon- and ion-mediated secondary processes. The multi-GEM (Gas Electron Multiplier [3]) multiplier could provide a solution: it has been recently shown that the triple- and quadruple-GEM structures can effectively operate in pure Ar and its mixtures with Ne and Xe, reaching rather high gains, up to $10^5$, at atmospheric pressure [4].

Another problem is that the density of noble gases near the boiling point, at normal pressure, is higher compared to that at room temperature. For example, in Ne, Ar and Xe the density difference is as large as a factor of 10.5, 2.8 and 1.6, correspondingly [5]. This means, that the operation of gas detectors at low temperature and atmospheric pressure can be equivalent to that at high pressure and room temperature. On the other hand, it was shown that the maximum GEM gain rapidly decreases with pressure in $Ar/CO_2$ and $Xe/CO_2$ [6].

In this paper we report on the performance of a triple-GEM detector in pure noble gases at high pressures, varying from 1 to 10 atm. The noble gases investigated are Ne, Ar and Xe. We show that the gain dependence on pressure is strongly affected by the gas nature.

The experimental setup is shown in Fig.1. 3 GEM foils (50 μm thick kapton, 80 μm diameter and 140 μm pitch holes, 28×28 $mm^2$ active area) and a printed curcuit board (PCB), mounted in cascade with 1.6 mm gaps, were installed within a stainless steel vessel. The vessel was filled with Ne, Ar or Xe at a certain pressure. The noble gases purity was 99.99%. The detector was irradiated with an X-ray tube through an Al window.

The GEM and PCB electrodes were connected to a resistive high-voltage divider, as shown in Fig.1. The divider was optimized in such a way as to maximize the gain in Ar at 1 atm and at the same time to prevent the parallel-plate amplification mode in inter-GEM (transfer) and GEM-PCB (induction) gaps. In particular, the voltage drops across GEMs were not equal and increased from the first to last GEM, similar to that used in [4]. Typical electric fields in the transfer and induction gaps, at 1 atm, were below 1.2, 3.0 and 2.8 kV/cm in Ne, Ar and Xe correspondingly. The same divider was used in the measurements with other pressures and gases. It should be remarked, however, that the optimized divider for them might be different.

The anode signal was readout from the PCB either in a current or pulse-counting mode. The anode current value was always kept below 100 nA, using X-ray attenuation filters, to prevent charging-up effects. The ratio of the anode current to the current recorded in the drift gap provides the gain value. The maximum attainable gain was defined as that at which neither dark currents nor anode current instabilities (discharges) were observed for at least about 1 min.

Fig.2 shows the gain-voltage characteristics of the triple-GEM detector in Ar, at different pressures. One can see that there are two types of the gain dependence on pressure. Below 3 atm, the maximum gain weakly depends on pressure, varying from $4\times10^4$ to $10^5$. In this pressure range it was limited by the onset of the dark current, of the order of few hundreds nA, most probably arisen due to the ion feedback between GEM elements [4]. At higher pressures, the maximum gain rapidly dropped down to below 10 at 7 atm. Here the limitation on the maximum gain was imposed by GEM discharges.

In Xe, the pressure dependence of the maximum gain also consisted of two parts: a slow decrease below 2 atm and very fast drop at higher pressures (Fig.3). On the other hand, the maximum operation gain was lower: it did not exceed $10^4$. The maximum gain in Xe was limited by discharges in the whole pressure range. In addition, among other gases studied Xe was found to be the worst in terms of the discharge detrimental effect: at least in two cases all 3 GEMs were completely destroyed after few discharges when operated in Xe, while in Ne and Ar even hundreds discharges did not result in noticeable degradation of the triple-GEM structure.

It is interesting, that the maximum gain (discharge) boundary in Ar and Xe at higher pressures looks like a barrier in the voltage drop across a GEM, of about 700 V, which cannot be overcome. This is probably related to the properties of the discharge mechanism in given gases.

Neon showed quite different behavior compared to Ar and Xe (see Fig.4). Unlike Ar and Xe, the maximum gain in Ne turned out to be a growing function of the pressure: it increased from $10^3$ at 1 atm to above $10^5$ at 7 atm. The limitation on the maximum gain in Ne was imposed by discharges. Note that the operation voltages in Ne are considerably lower compared to those in Ar and Xe. Another interesting observation is that the gain-voltage characteristics in Ne almost do not change with pressure, for above 5 atm, in contrast to Ar and Xe. This is unusual for traditional gaseous devices, for which one would rather expect the E/p behavior of the detector characteristics. The detector performance in Ne was studied in a pulse-counting mode as well, using a charge-sensitive amplifier: the data were in coherence with those obtained in the current mode.

We do not aware at the moment of any consistent explanation of GEM behavior at high pressures. We can only speculate that the violation of E/p scaling in Ne could indicate on the existence of some geometrical factors governing the gas amplification mechanism at high pressures, similar to that of the avalanche confinement in GEM holes considered in [8]. We also believe that the rather low cross-section of electron-atomic collisions in Ne, as compared to other gases [7], may play an important role.

In conclusion, we have studied for the first time the high-pressure operation of a triple-GEM detector in pure Ne, Ar and Xe. Neon showed quite different pressure dependence of the maximum gain as compared to Ar and Xe: in Ar and Xe the maximum gain drastically drops down for pressures exceeding 3 atm, while in Ne it increases with pressure up to 7 atm. In all the gases studied there exist an optimal pressure at which the triple-GEM detector has the maximum gain: $10^4$ at 1 atm, $10^5$ at 3 atm and $10^5$ at 7 atm in Xe, Ar and Ne correspondingly. One can see that the optimal pressures are close to those corresponding to appropriate gas densities near the boiling points. This means that the triple-GEM detector, in terms of gain characteristics, is a good candidate for the proposed cryogenic double-phase particle detectors. At the same time, the gas amplification mechanism in GEM at high pressures is still unclear. Further investigations are required.

We thank Drs. M. Leltchouk and D. Tovey for useful discussions.

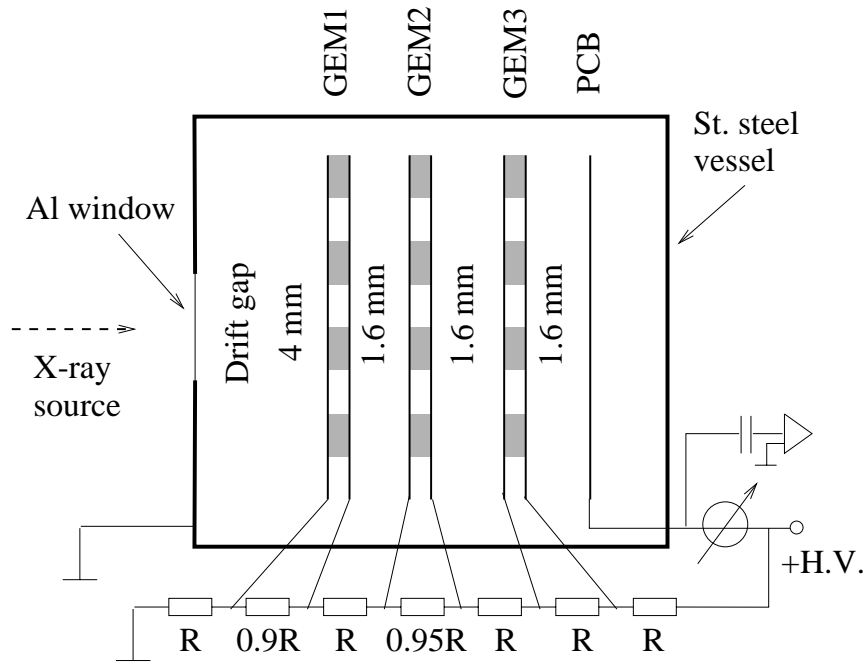

Fig.1 A schematic view of the triple-GEM detector operated at high pressures.

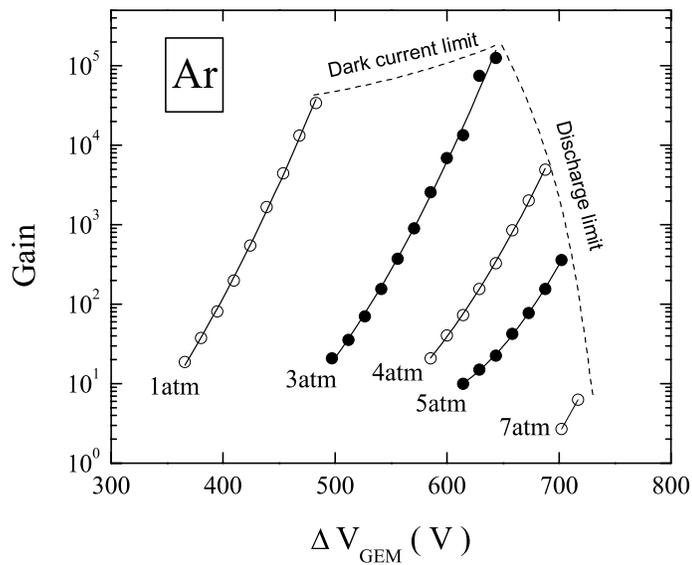

Fig.2 Detector gain as a function of the voltage drop across the last GEM, in Ar at different pressures.

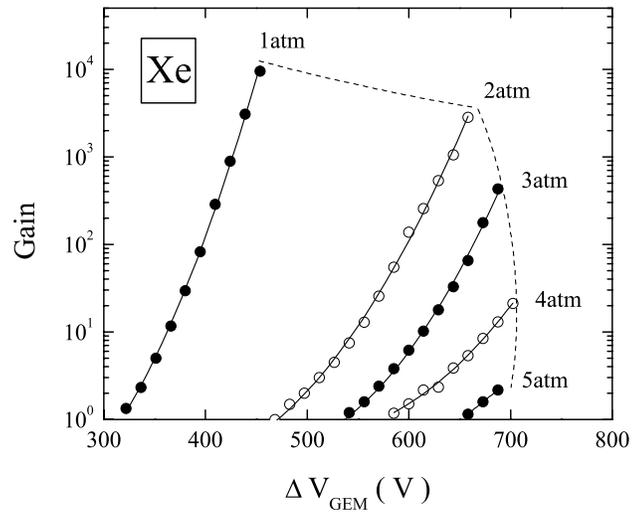

Fig.3 Detector gain as function of the voltage drop across the last GEM, in Xe at different pressures.

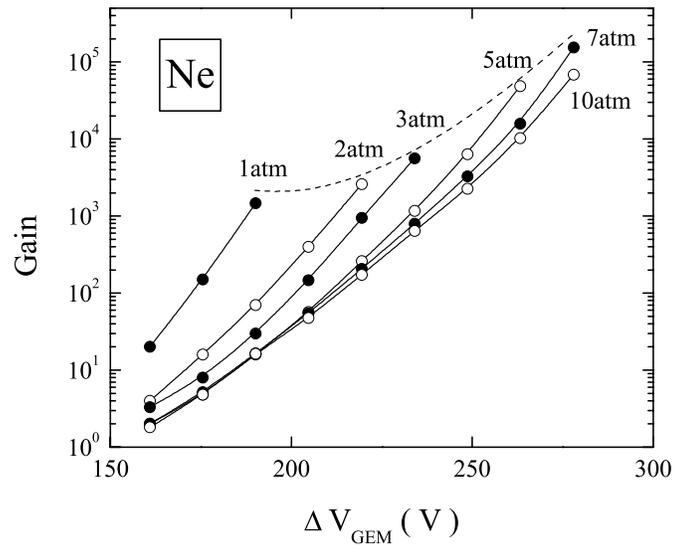

Fig.4 Detector gain as a function of the voltage drop across the last GEM, in Ne at different pressures.